# Transverse Photo-Voltage Induced by Circularly Polarized Light

Takafumi Hatano （畑野敬史）[1], Teruya Ishihara （石原照也）[1,2]*,

Sergei G. Tikhodeev[3] and Nikolay A. Gippius[3,4]

[1]Department of Physics, Tohoku University, Sendai, Japan

[2]Frontier Research System, RIKEN, Wako, Japan

[3]A. M. Prokhorov General Physics Institute, RAS, Moscow, Russia

[4]LASMEA, UMR 6602, Universite Blaise Pascal, Aubiere, France

*Corresponding author: t-ishihara@mail.tains.tohoku.ac.jp

We discovered that when circularly polarized light is obliquely incident on a two-dimensional metallic photonic crystal slabs, electrical voltage is induced perpendicularly to the incident plane. Signal sign is reversed by changing the sense of polarization or incident angle. The origin of this transverse photo-induced voltage is explained in terms of the force linear in light intensity induced by the asymmetry brought by angular momentum of incident light.

*Introduction* - Circularly polarized light beam has angular momentum. It has various effects on the optical response of matter and has been investigated over a long period [1]. Although optical angular momentum consists of spin and orbital contributions, we will consider only spin angular momentum due to circular polarization throughout this paper. Transfer of angular momentum of light to matter was theoretically discussed by Sadowsky [2] and Epstein [3] and experimentally demonstrated by Beth [4] in 1936, where he measured mechanical torque on a quarter wave plate given by circularly polarized light. Generation of circular motion of charged particles excited by circular polarization light has been extensively discussed in the field of plasma physics [5,6]. Recently magnetization induced by circularly polarized laser beam has been successfully demonstrated [7] and the discussions in the plasma physics is now linked to spintronics [8]. Another manifestation of the angular momentum is a lateral beam shift when circularly polarized light is incident on a flat surface, which is referred to as an Imbert-Fedorov effect [9]. Onoda et al. showed that it is largely enhanced in a photonic crystal and explained in terms of Berry's phase [10,11]. They also suggested strong analogy with the spin Hall effect. Recently Bliokh et al. discussed Coriolis effect in optics in the similar context [12,13]. The rotation-sense dependent phase shift was detected as a change of speckle pattern through multimode fiber and called optical Magnus effect [14]. These phenomena manifest themselves as small spatial shift. Although enhanced by artificial structures such as photonic crystals or long fibers, experimental demonstration of these effects is still very demanding.

While the angular momentum depends on the polarization, light always has translational momentum. Measurement of light momentum was carried out by Lebedev in 1900 with a radiometer setup [15]. After the advent of pulse lasers, light momentum can be conveniently detected as a voltage generated by the momentum of light. Photo-induced voltage (PIV) due to this mechanism is known as the photon drag effect, which was first discovered in semiconductors in 1970's [16]. Because of the extra complexity due to interband transition of careers in semiconductors, simple metals are preferred in order to investigate the photon drag itself. The theory of photon drag effect in metal was formulated in Ref. [17]. Ishihara reported PIV along a waveguide mode of dielectric grating on a thin metallic film [18]. Vengurlekar employed prism coupling configuration to demonstrate surface plasmon polariton (SPP) enhanced PIV on the surface of a metallic thin film [19]. PIV in photonic crystals with an asymmetric unit cell was investigated and readily explained in terms of the momentum conservation [20]. The strong diffraction of light in the metallic slab grating slab is responsible for transfer of momentum of light to free electrons. PIV observed in these experiments will be referred to as a longitudinal PIV (LPIV) in this Letter. Similarly we can expect that the transfer of angular momentum of light may be detected. From the symmetry consideration, the voltage, if any, should be observed perpendicularly to the incident plane, which can be referred to as a transverse PIV (TPIV).

The aim of this paper is to provide the very first experimental results on transverse voltage induced by *circularly polarized light* in the metallic slabs with artificial periodic structure. After describing the sample structure and fabrication method, angle resolved transmission spectra are shown to describe optical response due to SPP in our sample. Then transverse as well as longitudinal PIV are shown as a function of incident laser wavelength. The voltage is explained in terms of the second order electromagnetic force on free electrons in the metallic structure. Discussion based on symmetry will be given in order to elucidate the characteristic polarization dependence.

*Sample and experimental configuration-* In order to observe PIV, we specially designed our sample. First the metallic film has to be thin enough to ensure that the voltage generated near the surface of the film is not short circuited by the unilluminated part. In order to enhance electric field in the sample, we utilize SPP, to which laser excitation is coupled through the periodic modulation. Among various configurations of the periodic structure, hole array configuration was adopted because it provides strong modulation without worrying about disconnection. A 40 nm Au film was evaporated on a 1 cm$^2$ quartz substrate with a nominally 3 nm thick intermediate layer of Cr for better adhesion. A resist layer coated on the Au film was patterned by electron beam and the Au film was etched down to the substrate to form two dimensional periodic hole array as shown in the AFM image of Fig. 1(a). The structure is located at the center of the Au film, and has 1mm×1mm two dimensional square arrays of holes with the period of 500 nm. The hole diameter is 120 nm. To electrically isolate all four sides of the hole array structure, we cut around the structure as shown in Fig. 1(b). We measure the voltage across the structure on which we send the incident light with transverse and longitudinal configurations shown in Figs. 1(c) and (d).

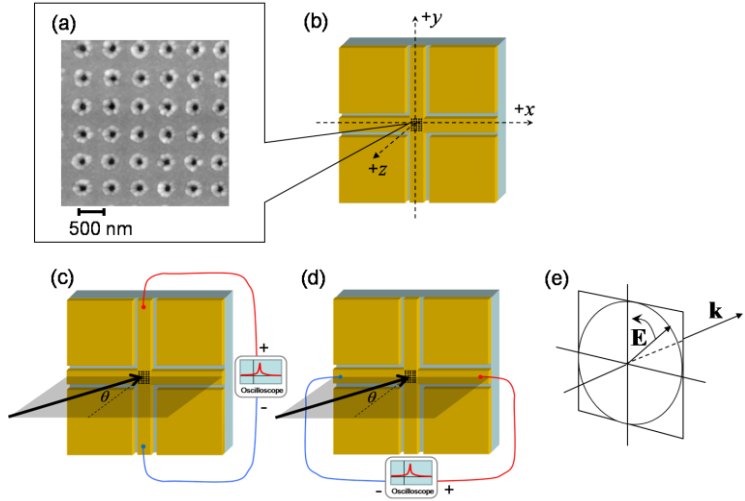

FIG. 1 (color online). (a) AFM image of our sample. (b) Axis definition. Experimental configuration of (c) TPIV and (d) LPIV. (e)Definition of left handed circular polarized light in this Letter.

Light from an optical parametric oscillator pumped by a frequency-tripled YAG laser is sent to the sample with the wavelength varying from 730 to 1200 nm. Pulse width and repetition rate of the laser are 5 nanoseconds and 10 Hz, respectively. The induced voltage is fed with $50\,\Omega$ input impedance to an oscilloscope triggered by Q-switch of the laser. There are two pairs of electrodes; for LPIV and TPIV measurements. When a pair of electrodes is connected to the oscilloscope, the other pair was kept open. Typical resistance of the sample was $20\,\Omega$. Measurement was made for left and right handed circular polarized light and s and p linear polarized light. To convert circularly polarized light from s-polarized light, achromatic quarter and half wave plates (Thorlab Inc.) are used. The sense of circular polarization is defined so that circular motion of electric field is observed at a fixed point toward the direction of propagation as shown in Fig. 1(e). For each laser wavelength, voltage corresponding to the laser pulse maximum is recorded to the computer.

*Experimental results-* To characterize optical response of our sample, we first measured angle resolved transmission spectra for p-polarized light. Figure 2(a) shows p-polarized transmission spectra for incidence angle from $\theta = 0$ deg to 39 deg in 3 degree steps. The sign of $\theta$ is defined so that $k_x$, the projection of wave vector to $x$ axis, is positive. There are two kinds of transmission dips which are ascribed to the excitation of SPP modes at air-metal and metal-substrate interfaces. The positions of dips shift to the longer wavelength side as the incident angle increases. At normal incidence a dip marked with a closed circle is located at 800 nm. It corresponds to -1st order SPP mode localized at the interface between the substrate and the Au. At slightly shorter wavelength from each dip, a kink corresponding to the onset of Bragg diffraction exists for lower angles. Another mode with closed squares corresponds to -1st order SPP at the interface between the air and the Au. Figure 2(b) shows corresponding transmission spectra for s-polarized light. We notice a SPP dip shifting to the shorter wavelength side as the incident angle increases. As a result for angles larger than 21 degree, only one SPP mode is excited around 1000 nm, which will be important for interpretation of the PIV spectra in our discussion.

Knowing the linear optical properties of our sample, we investigated PIV. For circularly polarized light, TPIV signal was observed near the SPP resonance. The signal intensity was of the order of 1mV for 2 MW/cm². Pulse width of observed voltage signal measured with the oscilloscope was 5 ns, which is following the incident laser pulse. In Fig.3(a), TPIV spectra at $\theta = 27$ deg is shown as a red solid curve for right circularly polarized light. At around 1000 nm where an SPP is excited, TPIV exhibits dispersive behavior. As for the signal for left-handed circular polarization, the sign reverses in the entire range of

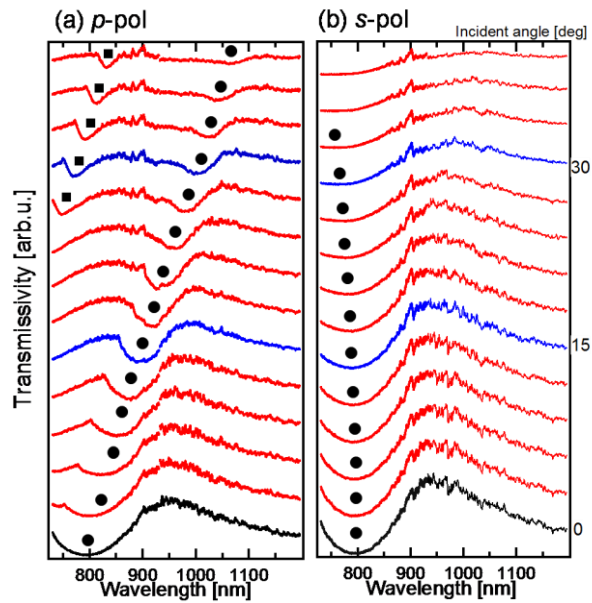

FIG. 2 (color online). (a) Transmission spectra for 0 to 39 deg with 3 deg step for p-polarized light. Closed circles: SPP mode excitation induced at the interface between metal and substrate. Closed squares: SPP excitation induced at the interface between metal and air. (b) Transmission spectra for s-polarized light.

the measurement as is shown as a blue dashed curve. For s- and p- polarization, no significant TPIV signal was observed, as shown by the orange dash-dotted and green dotted lines in Fig. 3 (a), respectively. In order to observe this polarization selection rule, fine adjustment of the sample orientation was required.

Figure 3(b) shows LPIV for the same sample at the same condition. At 770 and 980 nm, LPIV spectrum shows a peak corresponding to the excitation of SPP modes. The largest signal is observed for p-polarization with which SPP is excited. For s-polarization we do not see any feature at 980 nm. For left and right handed circular polarizations, LPIV signals, shown as red solid and blue dashed lines respectively, are identical, which is the average of signals for s- and p- polarization. Such polarization dependence will be readily understood in our later discussion.

For any polarization, TPIV (and also LPIV) is found to be an odd function of the incident angle. Accordingly no voltage is observed at the normal incidence (not shown). It is important to note that LPIV and TPIV are found to be proportional to the excitation intensity, which suggests that the effect is of the second order in the field amplitude.

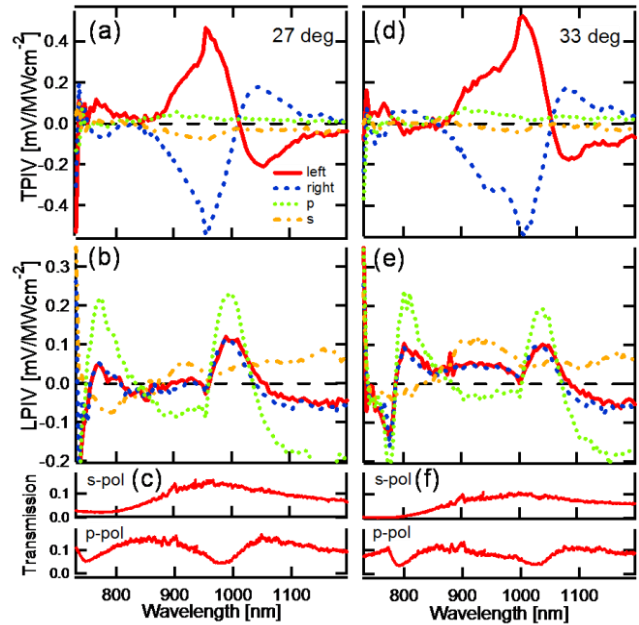

FIG.3 (color online). (a) TPIV and (b) LPIV spectra and (c) transmission spectra for s- and p-polarized light for incident angle of 27 deg. (d-f) Corresponding spectra for incident angle of 33 deg.

In Fig. 3(d-f), the corresponding spectra are shown for the incident angle of 33 degree. As the angle is increased, the prominent dispersive feature in TPIV spectra and the peak in LPIV spectra are both displaced as a whole to the longer wavelength side. They correspond to the dips at about 1000 nm (c) and 1030 nm (f) in p-polarized transmission spectra. As there is no dip in s-polarized transmission, it is clear that only one mode is involved in this case. On the other hand, at 790 nm, modes can be excited both for p- and s- polarization. As a consequence the spectral shape for LPIV and TPIV are more complicated. We will analyze the spectral response in the simpler situation later in the discussion.

*Discussion* - The electromagnetic DC force exerted on electrons in the perforated metallic film is calculated from the Lorentz force by treating $|\mathbf{E}|^2$ as perturbation and retaining the lowest order [21,22].

$$\vec{F}^{(2)}(r) = \text{Re}\left[\alpha[(\vec{E}\cdot\vec{\nabla})\vec{E}^* + \vec{E}\times(\vec{\nabla}\times\vec{E}^*)]\right]$$
$$= \text{Re}[\alpha]\vec{\nabla}|\vec{E}|^2 + \text{Im}[\alpha]\text{Im}\sum E_i\vec{\nabla}E^*_i \quad (1)$$

where $\alpha = e^2(i\gamma-\omega)/2m\omega(\omega^2+\gamma^2)$ and $\gamma$ is a damping constant. TPIV and LPIV are obtained by integrating *y* and *x* component of the force (1) along *x,y* and *z* directions. The *y* component of the integral corresponds to transverse PIV whereas the *x* component gives longitudinal PIV. Here, the first term of equation (1) can be neglected after integration because it is a conservative force. The second term of equation (1) can be written as

$$\text{Im}[\alpha]\text{Im}\sum E_i\vec{\nabla}E^*_i = A\vec{\eta}(\lambda,\theta),$$

where $A$ and $\vec{\eta}$ are defined as follows

$$A = \text{Im}[\alpha]|E_0|^2 k_0$$
$$\vec{\eta}(\lambda,\theta) = \text{Im}\left[\sum_{i=x,y,z}\frac{E_i}{E_0}\frac{\vec{\nabla}}{k_0}\frac{E^*_i}{E_0}\right]$$

where $A \sim 2 \times 10^{-15}$ dyn for incident light intensity of $I = 1$ MW/cm². $\vec{\eta}(\lambda,\theta)$ is a dimensionless vector characterizing the photo-response of the structure. Although our sample has square symmetry, the mirror symmetry through *y-z* plane is broken for the electromagnetic field due to the oblique incidence in our measurement. On the other hand the mirror symmetry for z-x plane is retained for any incident angle. Electromagnetic field in the sample excited by p-polarized light $\vec{E}_\mathbf{p}$ is classified according to the reflection operation to the mirror plane: $E_{py}$ and $E_{pz}$ are odd, while $E_{px}$ is even. Similarly for s-polarized light, electromagnetic field in the sample $\vec{E}_\mathbf{s}$ is classified: $E_{sy}$ is odd, while $E_{sx}$ and $E_{sz}$ are even. As a result $F_y^{(2)}$ is found to be odd. Therefore TPIV vanishes for s- and p-polarization. As for the circularly polarized light, however, the situation is different. For right (left) polarized light, the field in the sample is given as a superposition of the two fields:

$$\vec{E} = \frac{1}{\sqrt{2}}(\vec{E}_\mathbf{p} \pm i\vec{E}_\mathbf{s})$$

In this case, $F_y^{(2)}(r)$ includes even function of $y$,

$$\sum_{i=xyz} E_{pi}\nabla_y E_{si}^* \mp E_{si}\nabla_y E_{pi}^*$$

which gives rise to finite transverse voltage. Furthermore this expression explains that TPIV at the SPP resonance has dispersive spectral response, because the internal electric field changes its sign as the wavelength changes across the resonance. Since the expression contains only one $\vec{E}_\mathbf{p}$ for each term, it changes its sign at the resonance.

Now that the formula is known, it is tempting to numerically calculate the longitudinal and transverse response of our structure and to compare it with the experiment. Actually we have calculated the angle dependent transmission spectra for our structure with a scattering matrix method [23], resulting in a reasonable agreement with the experiment. PIV spectra, however, turned out to be very sensitive to the details of the field distribution inside the metallic structure. Thus we cannot make conclusive comparison of the data with the numerical calculation at the moment. It is highly demanding to calculate field distribution in two-dimensional metallic structure but will be extremely rewarding because various aspects of geometrical phase will be directly connected to the transverse voltage which is accessible in the experiment.

The transverse photo-voltage induced by the circularly polarized light could be referred to as "optical

Hall effect", although the name is already used to express the transverse shift of a light beam [10]. The voltage we discovered might be interpreted as the counteraction of the beam shift to the free electrons in the metal.

We would like to thank Prof. Naoshi Nishimura of Kyoto University and Prof. Masaru Onoda of Akita University for fruitful discussions and suggestions from theoretical point of view. This research was partially supported by the Ministry of Education, Science, Sports and Culture, Grant-in-Aid for Scientific Research (B), 17340092 and ANR Chair of Excellence Program, the Russian Academy of Sciences, and the Russian Foundation for Basic Research.